\documentclass[final,5p,twocolumn]{elsarticle}
\usepackage{graphicx}
\newcommand{\eq}{\begin{equation}}
\newcommand{\eqx}{\end{equation}}
\newcommand{\eqn}{\begin{eqnarray}}
\newcommand{\bi}{\begin{itemize}}
\newcommand{\eqnx}{\end{eqnarray}}
\newcommand{\ei}{\end{itemize}}
\newcounter{hran}

\journal{Physics Letters B}

\begin{document}

\begin{frontmatter}

\title{Minimal Inflation}

\author[lag]{Luis \'Alvarez-Gaum\'e}
\ead{Luis.Alvarez-Gaume@cern.ch}
\address[lag]{Theory Group, Physics Department, CERN, CH-1211, Geneva 23, Switzerland.}

\author[cg,lag]{C\'esar G\'omez}
\ead{cesar.gomez@uam.es}
\address[cg]{Instituto de Fisica Teorica UAM/CSIC, Universidad Autonoma de Madrid, E-28049 Madrid, Spain.}

\author[rj,lag]{Raul Jimenez}
\ead{jimenez@icc.ub.edu}
\address[rj]{ICREA \& Institute of Sciences of the Cosmos (ICC), University of Barcelona, 08028 Barcelona, Spain.}

\begin{abstract}
Using the universal $X$ superfield that measures in the UV the violation of conformal invariance we build up a model of multifield inflation. The underlying dynamics is the one controlling the natural flow of this field in the IR to the Goldstino superfield once SUSY is broken. We show that flat directions satisfying the slow roll conditions exist only if R-symmetry is broken. Naturalness of our model leads to scales of SUSY breaking of the order of $10^{11-13}$ Gev, a nearly scale-invariant spectrum of the initial perturbations and negligible gravitational waves. We obtain that the inflaton field is lighter than the
gravitino by an amount determined by the slow roll parameter $\eta$.
The existence of slow-roll
conditions is directly linked to the values of supersymmetry and R-symmetry breaking
scales. We make cosmological predictions of our model and compare them to 
current data.
\end{abstract}


\end{frontmatter}

\section{Introduction}

In spite of the enormous success of inflationary cosmology \cite{guth:1981,mukhanov:1981,
sato:1981,albrecht.steinhardt:1982,guth/pi:1982,hawking:1982,linde:1982,starobinsky:1982,
bardeen/steinhardt/turner:1983} at describing the observed properties of the Universe, 
we are still missing a derivation from first principles where the inflaton field is 
identified with one, or several, fundamental fields in particle physics. This manifests itself the in 
the fact that we still do not count with a natural way of identifying the inflaton field
and the properties of its potential required to satisfy experimental constraints
\cite{liciaps,wmap5}. 

It was quickly realized after the inflationary scenario was proposed more than 30 years ago, that supersymmetry could provide a natural scenario with plenty of flat
directions which could lead to inflation \cite{ellisetal, ovrutetal, rossetal, dvalietal, randalletal, lythriotto}.  When the theory couples to supergravity, there are a number of new
problems that appear \cite{copelandetal:1993}, and we will discuss some of them
later on.

Current observational constraints from CMB temperature and polarization experiments
and large-scale structure limit the amount the inflaton field has moved to 
approximately $< 2 M_{pl}$\cite{raulcmb}, where $M_{pl}$ is the reduced Planck 
mass. Therefore, inflationary models that search for the inflaton at very large 
energies, like for example chaotic inflation,  are severely constrained already by 
current observations. With the current  new generation of CMB experiments (Planck, 
EBEX, Spider, SPUDS etc...) it will be possible to further constraint how much the 
inflaton field has displaced during the inflationary period that gave rise to our 
current casual horizon.  It is therefore useful to revisit again the problem of 
steep directions in SUGRA models to understand if a flat direction can be obtained 
at all.

In this paper we will suggest a natural embedding of inflationary dynamics
in the effective low-energy Lagrangian describing supersymmetry breaking.  Our
approach will be quite independent of the microphysics underlying supersymmetry
breaking, and will only rely on universal properties of this symmetry.  Since
we are not committing ourselves to any particular microscopic realization
of supersymmetry breaking, some of our comments about reheating
for instance will be rather sketchy.  A more detailed and precise presentations
of our ideas will appear elsewhere \cite{usindetail}.  Like most inflationary
theories containing supersymmetry, we present a simple model of multifield
inflation (sometimes called hybrid) \cite{lindehybrid}, 
identify naturally the inflaton field and its
potential, and then fit a few observational data to estimate the few parameters
of our model.  We compute, in particular, the number of e-folding and the amplitude
of density fluctuations at horizon crossing.  It is surprising to find that
the scale of supersymmetry breaking indicated by this analysis is 
between $10^{11}-10^{14}$GeV.  An interesting spin-off of our model is that
the inflaton is lighter than the gravitino by an amount $\sqrt{\eta}$,
where $\eta$ is one of the slow roll parameters (see below).

We would like to stress that in this paper we are always assuming F-breaking of
supersymmetry.  In D-breaking scenarios our arguments do not apply, at
least as presented here \footnote{We thank Gia Dvali for raising this
point.  See for instance the last entry in \cite{dvalietal}}.

\section{General framework}

Supersymmetry is a natural framework to define inflationary scenarios for 
two main reasons.  First of all, SUSY naturally leads to the existence of 
flat, or nearly flat directions (pseudomoduli), allowing for slow roll scenarios.
Second, and more important, the order parameter of supersymmetry breaking
is the vacuum energy density.  Hence, naturally associated with its breaking,
supersymmetry contains two main ingredients necessary in inflationary scenarios:
vacuum energy and reasonably flat directions. 

In a remarkable recent work, Komargodski and Seiberg \cite{komarseiberg}
have presented a new formalism to understand  supersymmetry breaking, its
general properties, its non-linear realizations \cite{volkovakulov}, and
a systematic way to understand the low-energy couplings of goldstinos
to other fields.  Although many things were known before (see references
in \cite{komarseiberg}) this work, the presentation is quite insightful,
and it played a major part in the inspiration of this work. 

The basic starting point in \cite{komarseiberg} is
the Ferrara-Zumino multiplets of currents \cite{ferrarazumino}.  A vector
superfield composed of the R-symmetry current, the supercurrent, and the energy
momentum tensor. This vector superfield satisfies the general relation:
\eq
\label{ferrara}
\bar D^{\dot{\alpha}}J_{\alpha,\dot{\alpha}}=D_{\alpha}\,X.
\eqx
The chiral superfield $X$ is essentially defined uniquely \footnote{The ambiguities 
in the supercurrent multiplet and $X$ are related to 
improvement terms in the various currents.}
in the ultraviolet.  Following \cite{komarseiberg} the superfield $X$ 
has the following properties:
\begin{itemize}
\item
In the UV description of the theory, it appears in the right hand side of
\ref{ferrara}, where it represents a measure of the violation of conformal
invariance.
\item
The expectation value of its $\theta^2$ component is the order parameter
of supersymmetry breaking.  In this work we are only considering $F$-breaking
of supersymmetry. We denote by $f$ the expectation value of the $F$-component
of $X$.  It will sometimes be useful to write $f\,=\,\mu^2$, where $\mu$
is the microscopic scale of supersymmetry breaking. 
\item
When supersymmetry is spontaneously broken, we can follow the flow of $X$ 
to the infrared (IR).  In the IR this field satisfies a non-linear constraint
and becomes the ``goldstino" superfield \footnote{A modified version of the
nonlinear constraint (\ref{xsquare}) appears when one considers spontaneous
R-symmetry breaking.  In that case, the goldstino and the corresponding axion
will be part of the same multiplet.}.
\eq
\label{xsquare}
X_{NL}^2\,=\,0,
\eqx
\eq
\label{goldstinosuper}
X_{NL}\,=\, { G^2\over 2\,F}\,+\,\sqrt{2}\,\theta\,G\,+\,\theta^2\, F.
\eqx
The scalar component $x$ of $X$ becomes a goldstino bilinear.  Its fermionic
component is the goldstino fermion $G$, and $F$ is the auxiliary field that
gets the vacuum expectation value.  A major part in the analysis in 
\cite{komarseiberg} is based on this novel nonlinear constraint satisfied by
the superfield $X$ in the IR. As shown there, the correct normalization
of the goldstino superfield to derive all relevant low-energy theorems of broken
supersymmetry is $X_{NL}={3\over 8 f} X$.
\item
Finally, $X$ generalizes the usual spurion couplings appearing in the description
of low-energy supersymmetric lagrangians. If $m_{soft}$ describes the soft supersymmetry
breaking masses at low energies, the standard spurion in the lagrangian is replaced
by ${m_{soft}\over f} X_{NL}$.  This allows one to write the leading low-energy
couplings of the goldtino to other matter fields.  
\end{itemize}
Since we are going to consider goldstino couplings, we will work with a field whose
expectation values are well below the Planck scale.  

{\bf Our proposal} is to identify in the UV the inflaton field with the scalar component
of the superfield $X$.  Since $X$ is defined uniquely (up to the ambiguity mentioned
in footnote one) in the UV, this provides a well defined prescription.  Furthermore,
we will identify the inflationary period precisely with the flow of $X$ from the UV to 
the IR i.e. $X\rightarrow\, X_{NL}$. Note that by making this assumption we do not 
need to think  of the inflaton as any extra fundamental field. In fact, independently 
of how SUSY  is broken, and what is the underlying fundamental theory 
we can always identify  the $X-$superfield as well as its scalar 
component $x$. More importantly, by making 
this assumption we are identifying the vacuum energy driven 
inflation with the actual SUSY breaking order parameter.

In the supergravity context, once we have the K\"ahler potential $K(X,{\bar X})$ and
the superpotential $W(X)$, the full scalar potential is given by \cite{ferrarapotential}:
\eq
V= e^{\frac{K}{M^2}}(K_{X, \bar X}^{-1} DW \bar DW -\frac{3}{M^2}|W|^2)
\eqx
with
\eq
D\,W\,=\,\partial_X\,W\,+\,{1\over M^2} \partial_X K\, W.
\eqx
$M$ is the high energy scale below which we can write the effective action
describing the dynamics of the $X$-superfield. It could be the Planck scale,
or a GUT scale depending on the microscopic theory.  We will work well below
the scale $M$, and for simplicity take $M=M_{pl}$
In equation (4) we can see one of the basic problems in
supergravity inflation \cite{copelandetal:1993}.  As we will see later on,
to satisfy the slow roll conditions, a necessary condition is that the
$\eta$-parameter, defined by:
\eq
\eta \,=\, M^2_{pl} \frac{V''}{V},
\eqx
be much smaller than one.  If we choose a K\"ahler potential $K(X,{\bar X})$
with R-symmetry, for instance the canonical one $K(X,{\bar X})=X{\bar X}+\ldots$,
where the $\ldots$ represents a function of $X{\bar X}$, it is easy to see
that from the exponent of (4) we always get a contribution to $\eta$ equal
to $1$: $\eta=1+\ldots$, no matter which component of $X$ is taken as the
inflaton field.  This of course violates the slow roll conditions.  Since
we are considering a situation with supersymmetry breaking and gravity (early
universe), we cannot exclude supergravity from the picture, and this leads
to the $\eta$-problem in these theories. 

The simplest way out of this problem without unreasonable  fine tuning, is to
have explicit R-symmetry breaking in the K\"ahler potential \footnote{R-symmetry
is a well-known problem in phenological applications of supersymmetry.  R-symmetry
does not allow soft breaking masses for the gauginos; and spontaneous breaking
of the symmetry may lead to axions with unacceptable couplings. Often one wants to
preserve R-parity to avoid other possible phenomenological disasters.}.  If we
have explicit R-breaking, the expansion of $V$ for small fields takes the form:
\begin{eqnarray}
X & = & M (\alpha\,+\,i\,\beta) \\
V & = & f^2 (1\,+\,A_1 (\alpha^2+\beta^2)+ B_1(\alpha^2-\beta^2)+\ldots)
\end{eqnarray}
$f$ is the supersymmetry breaking parameter representing the expectation value
of an $F$-term, and hence with square mass dimensions.  We assume that $V$
is locally stable at least during inflation.  Hence $A_1\pm B_1 > 0$.  We 
express the potential in terms of the dimensionless fields $\alpha,\beta$.
Their masses can be read off from (8):
\eq
m_{\alpha}^2\,=\,{2\,f^2\over M^2}(A_1+B_1),\qquad m_{\beta}^2\,=\,{2\,f^2\over M^2}(A_1-B_1).
\eqx
The numbers $A_1,B_1$ are taken to be $O(1)$.

One could be more explicit, and choose some supersymmetry breaking superpotential, 
like $W=f X$, and K\"ahler potential explicitly breaking R-symmetry, like:
$K=X{\bar X}+(c/M^2)(X^3 {\bar X}+ X {\bar X}^3)+\ldots$ as in \cite{komarseiberg}
leading to an effective action description of $X$ for scales well below $M$.
At this stage, we prefer not to consider explicit examples of UV-completions of
the theory.

We consider the beginning of inflation well below $M$, hence the initial conditions
are such that $\alpha,\beta <<1$.  In fact, since $\beta$ is the lighter field,
we take this one to be the inflaton, and consider that initially $\alpha,\beta \sim
\sqrt{f}/M$.  For us the inflationary period goes from this scale until the
value of the field is close to the typical soft breaking scale of the problem
$m_{soft}$, where the field $X\rightarrow X_{NL}$  (\ref{xsquare}), at this
scale $X_{NL}$ behaves like a spurion \cite{komarseiberg} and as shown in Ref. \cite{komarseiberg}, 
the leading couplings to low-energy supersymmetric matter
can be computed as spurion couplings, for instance \footnote{The details
can be found in\cite{komarseiberg} section 4, in particular around equations (4.3,4).},
if $Q,V$ represent respectively low energy chiral and vector superfields, we can
have the couplings:
\begin{eqnarray}
{\cal L}\,&=&\,-\int\,d^4\theta\,\left|{X_{NL}\over f}\right|^2 m^2 Q e^V {\bar Q}\, \\ \nonumber
& + & \int d^2\theta\,{X_{NL}\over f}\left({1\over 2} B_{ij} Q^i Q^j+\ldots + h.c.\right)
\end{eqnarray}
plus gauge couplings. 

Once we reach the end of inflation, the field $X$ becomes nonlinear, its scalar
component is a goldstino bilinear and the period of reheating begins. The details
of reheating depend very much on the microscopic
model.  At this stage one should provide details of the ``waterfall" that turns
the huge amount of energy $f^2$ into low energy particles.  Part of this energy
will be depleted and converted into low energy particles through the soft couplings
in (10), and hence we can in principle compute a lower bound on the reheating
temperature.  Before  making some comments on the reheating period, 
we analyze the cosmological
consequences of a potential as simple as (8), as well as the assumptions we
have made earlier about the inflaton and its range as inflation takes place. 

\section{The Inflaton Potential and Slow Roll Conditions}

To study the conditions under which our potential provides inflation consistent with the
latest cosmological constraints, we examine the slow-roll parameters, defined as \cite{LiddleLyth92}:
\eq
\epsilon = \frac{M^2_{pl}}{2} \left(\frac{V'}{V}\right)^2,
\eqx
\eq
\eta = M^2_{pl} \frac{V''}{V},
\eqx
where $M_{pl}$ is the reduced Planck mass and ' denotes derivative with respect to the 
inflaton field. The observables are then expressed in terms of the above slow roll 
parameters as:

\begin{eqnarray}
n_S &=& 1 - 6 \epsilon + 2 \eta, \\
r &=& 16 \epsilon \\
n_t &=& - 2 \epsilon, \\
\Delta^2_R &=& \frac{V M_{pl}^4}{24 \pi^2 \epsilon}.
\end{eqnarray}
$n_S$ is the slope of the scalar primordial power spectrum, $n_t$ is the corresponding tensor
one, $r$ is the scalar to tensor ratio and $\Delta^2_R$ is the amplitude of the initial 
perturbations. All these numbers are constrained by current 
cosmological observations \cite{liciaps, wmap5, liciaps2}. We will use their constraints to explore the 
naturalness of our inflationary trajectories. Inflation takes place when the slow-roll 
parameters are much smaller than $1$.

We will use the amplitude of initial perturbations and the number of efoldings
to fit some of the paramenters of the toy model in the previous section.  Recall
that the potential in the range of interest is:
\eq
V  =  f^2 (1\,+\,A_1 (\alpha^2+\beta^2)+ B_1(\alpha^2-\beta^2)+\ldots),
\eqx
which appears in figure 1.  We can compute $\epsilon, \eta$ while rolling in the $\beta$ direction:
\begin{eqnarray}
\epsilon & = & 2 \left( \,(A_1-B_1)\beta\, \right)^2+\ldots \\
\eta & = & 2\, (A_1-B_1)+\ldots,
\end{eqnarray}
since $\beta <<1$, $\epsilon$ is naturally small.  We can make $\eta$ small
by a slight fine tuning of the difference $A_1-B_1$.  We will write $\eta$
later as a ratio of the inflaton and gravitino masses.  Once the slow roll
conditions are satisfied, we can compute the number of efoldings (see
for instance \cite{KolbTurner, Mu}):
\eq
N\,=\,{1\over M} \left|\int\,{d x\over \sqrt{2 \epsilon}}\right|\,=\,
\left|\int_{\beta_i}^{\beta_f} {d\beta\over 2\sqrt{\epsilon}}\right|
\eqx
From (19) we get:
\eq
N\,=\,{1\over \sqrt{2} |A_1-B_1|}\left|\,\log {\beta_f\over \beta_i}\right|.
\eqx
In most models of supersymmetry breaking, the gravitino mass is given by:
\eq
m_{3/2}\,=\,{f\over M},
\eqx
hence, we can rewrite the parameters and masses in (9) as:
\eq
\left|A_1-B_1\right|\,=\,{1\over 2}{m_{\beta}^2\over m_{3/2}^2},\qquad
\left|A_1+B_1\right|\,=\,{1\over 2}{m_{\alpha}^2\over m_{3/2}^2},
\eqx
thus:
\eq
N\,=\,\sqrt{2} \left({m_{3/2}\over m_{\beta}}\right)^2
\left|\,\log {\beta_f\over \beta_i}\right|
\eqx
The number of efoldings is considered normally to be between $50-100$.
Finally we will use the amplitude of initial perturbations 
to get one extra condition in the parameters of our potential.  Using
\cite{wmap5} (16) can be written as:
\eq
\left({V\over \epsilon}\right)^{1/4}\,=\,{f^{1/2}\over 2^{1/4}\left(
|A_1-B_1|\beta\right)^{1/2}}\,=\,.027 M,
\eqx
where $\beta$ is taken at $N$-efoldings before the end of inflation.
Summarizing, the two cosmological constraints we get on the parameters
of our potential can be written as:
\eq
N\,=\,\sqrt{2} \left({m_{3/2}\over m_{\beta}}\right)^2
\left|\,\log {\beta_f\over \beta_i}\right|,
\eqx
\eq
2^{1/4}\,{m_{3/2}\over m_{\beta}}\,\left( {\sqrt{f}\over M}\right)^{1/2}\,=\,0.027,
\eqx
and the $\eta$ parameter can be written as:
\eq
\eta\,=\left({m_{\beta}\over m_{3/2}}\right)^2.
\eqx
We take $\beta_i$ above the supersymmetry breaking scale $\sqrt{f}/M=\mu/M$,
and $\beta_f$ close to $m_{soft}/M$, therefore we can easily get values for $N$ between
$50-100$ for moderate values of $\eta$, which is expressed here as the square
of the ratio of the inflaton to the gravitino mass. It is interesting to notice
that from (27), we can write the supersymmetry breaking 
scale $\mu$ in terms of the $\eta$-parameter:
\eq
{\mu\over M}\,\approx \, 5.2\,\, 10^{-4}\,\, \eta.
\eqx
Hence for a value of $\eta\sim .1$ we can get $\mu\sim 10^{13}$ GeV.  Lower
values of the supersymmetry breaking scale can be obtained by reducing $\eta$.
However, since the inflaton mass is
\eq
m_{\beta}\,=\,m_{3/2}\,\sqrt{{\eta}},
\eqx
we may end up with an inflaton whose mass is substantially lighter than the
gravitino.  For these values of $\eta,\mu$, we have that $\beta_i\sim 10^{13}/M, \beta_f\sim 10^3/M$, and the number of efoldings
is $\sim 110$.

We conclude then that with moderate values of $\eta$ between $.1-.01$ we can
get supersymmetry breaking scales between $10^{11}-10^{13}$ without major
fine tunings.  We easily get enough efoldings, and furthermore, the inflaton
is lighter than the gravitino by an amount given by $\sqrt{\eta}$.

For the above range of parameters we can compare the 
predicted value of $n_S$ in our model with observational constraints. This is shown in the right panel of Fig.~1. 
The yellow region is the current cosmological constraints from WMAP5 \cite{wmap5} and the other colored areas 
are the predictions for our model with minimal fine tuning for an stable (unstable) $X$ potential, i.e. the field 
is concave (convex) respectively. The constraints will improve greatly when the Planck satellite releases its results next year, and therefore our model can be tested much more accurately. 

Reheating can proceed in many ways, since we have not provided a detailed
microscopic model.  Once in the nonlinear regime, the $X_{NL}$ field (whose
scalar component is made of a goldstino bilinear) could efficiently convert
the $f^2$-energy density into radiation. 
We can calculate the amount of entropy and particle density by using the Boltzman equation and assuming that the pair of Goldstinos will have an out-of-equilibrium decay\cite{KolbTurner}. Using that

\eq
T_{RH} = 10^{-10} \left ( \sqrt f/GeV \right )^{3/2} GeV
\eqx
we obtain a range  $10^7 < T_{RH} < 10^9$. This produces a particle abundance of $n_{\chi} \sim 10^{70-90} $ which are standard values. We can also compute the amount of entropy generated by the out-of-equilibrium decay as 
\eq
S_f/S_i = 10^7 (\sqrt f/GeV)^{-1/2}
\eqx
which yields values in the range $10$ to $1$, and assures that there is no entropy overproduction.
We could also compute the depletion
of this energy through the soft couplings (10) yielding very similar values as above. In both cases, we can get sufficient reheating with temperatures between $\sqrt{f}$ and
a fraction of $m_{3/2}$.  The true value depends very much on the details
of the microscopic model. However, there seems to be no obstruction to
reheating the universe to and acceptable value of temperature, particle
abundances and entropy.  We are currently working in a more detailed theory
incorporating our scenario \cite{usindetail}.

\begin{figure*}
\begin{center}
\includegraphics[width=.8\columnwidth]{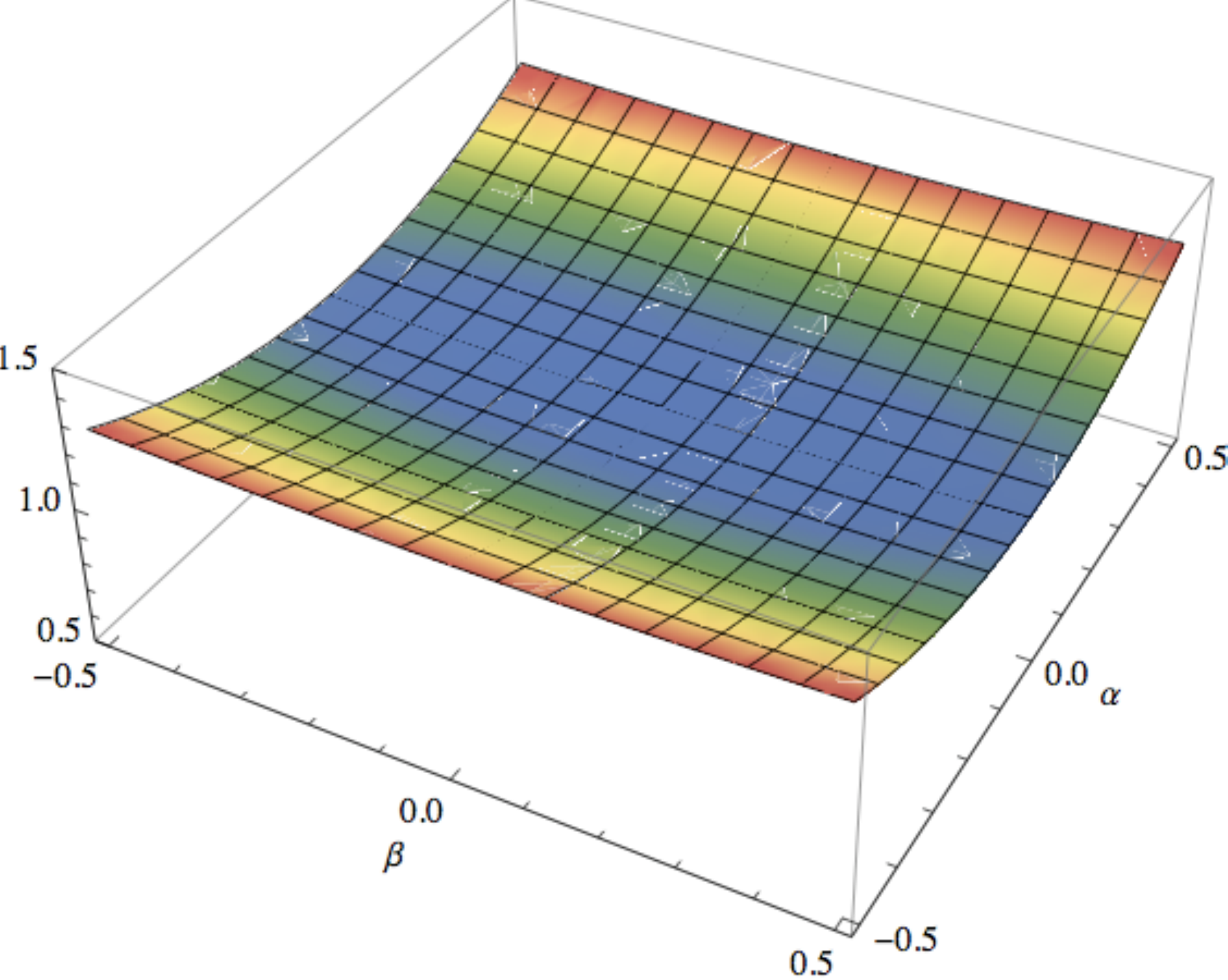}
\includegraphics[width=.8\columnwidth]{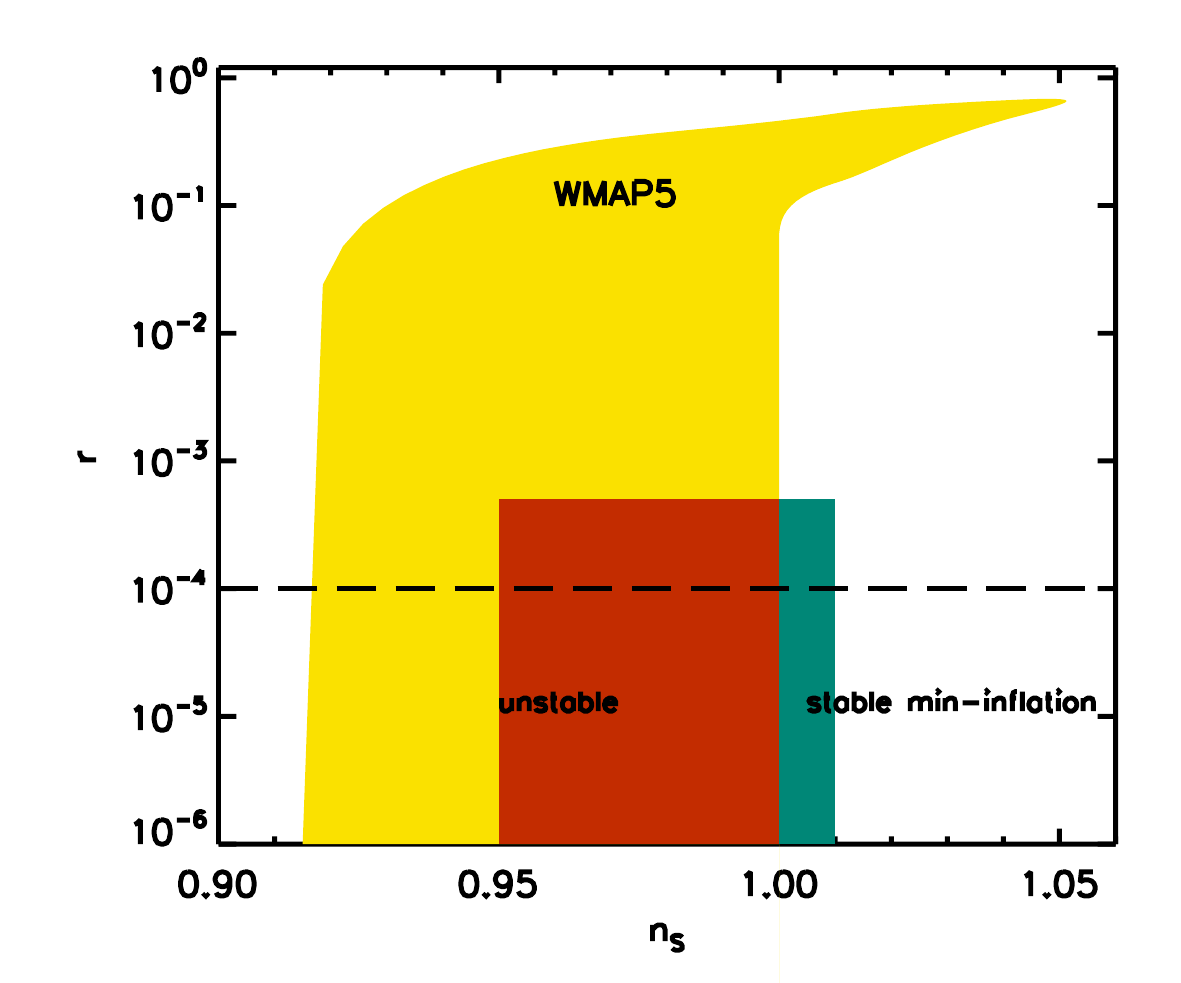}
\end{center}
\caption{Left panel: The potential as a function of  ($\alpha$) and ($\beta$) components of the field $X$. Note the nearly
flat direction ($\beta$) that we use for our inflationary trajectories.  Graceful exit and particle creation 
occurs in the non-linear part of the $X$ field.
Right panel: WMAP5 cosmological constraints (yellow region) in the $r-n_S$ plane. For no-fine-tuned minimal inflation models the green and red area show our predictions for both cases of a stable (concave) potential and unstable (convex) potential. The Planck satellite will be able to provide significantly tigther constraints on $r$ and especially $n_S$ (at the $< 0.5\%$ level) thus further constraining our model. The dashed line is the limit in $r$ that can be achieved with an ideal CMB polarization experiment \cite{raulcmb}}
\label{fig1}
\end{figure*}

\section{Conclusions}

In this short note we have studied the possibility of having supersymmetry
breaking as the driving force of inflation.  We have used the unique chiral
superfield $X$ which represents the breaking of conformal invariance in the UV,
and whose fermionic component becomes the goldstino at low energies.  Its 
auxiliary field is the $F$-term which gets the vacuum expectation value
breaking supersymmetry.

It is crucial in our analysis to have explicit R-symmetry breaking along
with supersymmetry breaking.  This allows us to avoid the $\eta$ problem
in supergravity and to take the supersymmetric limit.  The simplest model
we obtain describes the components of $X$ well below the Planck scale.
It is written in terms of three parameters:  the supersymmetry breaking
parameter $f$ and the masses of the real and imaginary components of the
field $x$ (the scalar component of X).  In our analysis the imaginary
part of $x$ plays the role of the inflaton, and its mass was shown to
be smaller than the gravitino mass by an amount given by $\sqrt{\eta}$.
This imaginary component represents a pseudo-goldstone boson, or rather,
a pseudomoduli.  In supersymmetric theories such fields abound, and any
of them could be used to construct some form of hybrid inflation.  In 
our case, however, we want to use the minimal choice that is naturally
provided by the universal superfield $X$ that must exist in any supersymmetric
theory.

Since we have not presented any detailed model, the cosmological consequences
are a bit rudimentary, especially concerning reheating at the end of inflation.
However, the comparison of the simplest model with present data, yields
very interesting values for the supersymmetry breaking scale, and the ratio
of the inflaton and gravitino masses.  These are bonuses which come
directly from the observations of the initial density perturbations 
from WMAP data \cite{wmap5}.  The fact that the inflaton is lighter
than the gravitino may have interesting low-energy phenomenological
implications.  Furthermore in this simple model
it is easy to obtain sufficient number of efoldings with moderate values
of the $\eta$ parameter.  

To explore our proposal in more detail, it is important to construct
an explicit model, even if not very realistic, in order to understand
in more detail the end of inflation, the reheating mechanisms, and
also the fine structure of the inflaton potential.  We hope to report
on this in the near future \cite{usindetail}.
\vskip.5cm
{\bf Acknowledgements}

We would like to thank G. Dvali, G. Giudice, J. Lesgourgues, S. Matarrese,
G. Ross, Nathan Seiberg, M.A. V\'azquez Mozo, and L. Verde for useful discussion.  C.G.
and R.J. would like to thank the CERN Theory Group for hospitality while
part of this work was done.


\end{document}